\documentstyle[12pt]{article} %  or [twocolumn,12pt]
\def\Acal{{\cal A}}
\def\Apara{A^\parallel}	\def\Aperp{A^\perp}	\def\Avperpl#1{A_v^{#1\perp +}}
\def\bareta{\bar\eta}			\def\bpsi{\bar\psi}
\def\bhvneg{\bar h_v^-}			\def\bhvpos{\bar h_v^+}
\def\D{{\cal D}}
\def\Dperp{D^\perp}			\def\Dperpsl{\rlap/\!D^\perp}
\def\Dslash{\rlap/\!D}
\def\del{\partial}
\def\det{{\rm det}}
\def\eq{\begin{equation}}	\def\endeq{\end{equation}}
				\def\endeql#1{\label{#1} \end{equation}}
\def\eqa{\begin{eqnarray}}	\def\endeqa{\end{eqnarray}}
				\def\endeqal#1{\label{#1} \end{eqnarray}}
\def\eqr#1{eq.~(\ref{#1})}	\def\Eqr#1{Eq.~(\ref{#1})}
\def\frac#1#2{{#1 \over #2}}
\def\Gcal{{\cal G}}
\def\Gcalperp{{\cal G^\perp}}
\def\Ham{{\cal H}}
\def\hc{{\rm h.c.}}
\def\hvneg{h_v^-}			\def\hvpos{h_v^+}
\def\journal#1#2#3#4{{\it #1} {\bf #2}, {#3} (19#4)}
\def\Lagr{{\cal L}}
\def\LHQEFT{{\cal L}_{\rm HQEFT}}
\def\LvHQ{L_{\rm HQ}^v}
\def\LmdQCD{\Lambda_{\rm QCD}}		\def\LmdSI{\Lambda_{\rm SI}}
\def\Pipara{\Pi^\parallel}		\def\Piperp{\Pi^\perp}
\def\Ppara{{\cal P}_{\mu\nu,v}^\parallel}
\def\Pperp{{\cal P}_{\mu\nu,v}^\perp}
\def\smlfrac#1#2{{\textstyle {#1 \over #2}}}
\def\vD{v\cdot D}
\def\vphiv{\varphi_v}
\def\vslash{\rlap/ v}
\def\Z{{\cal Z}}
\def\bzeta{\bar\zeta}

\newcounter{saveeq}
\def\alpheq{\setcounter{saveeq}{\value{equation}}%
\addtocounter{saveeq}{+1}
\setcounter{equation}{0}%
\renewcommand{\theequation}{\mbox{\arabic{saveeq}\alph{equation}}}}
\def\reseteq{\setcounter{equation}{\value{saveeq}}%
\renewcommand{\theequation}{\arabic{equation}}}

\begin{document}
\begin{titlepage}
\rightline{UCSD-TH-97-23}
\rightline{1997}
\vskip.5in
\begin{center}
 {\Large \bf Heavy Particle Effective Field Theories}
 \vskip.3in
 {{\bf Clarence L. Y. Lee}\footnote{\tt cl@ucsd.edu}\\
   \vskip.2cm {\it Department of Physics}\\
   \vskip.2cm {\it University of California, San Diego}\\
   \vskip.2cm {\it La Jolla, California 92093-0319}}
 \vskip.2cm
%PACS-96 numbers: 
\end{center}

\vskip.5in
\begin{abstract}

Starting from a theory of heavy particles and antiparticles, the path 
integral formulation of an effective field theory which describes the low 
momentum interactions is presented. The heavy degrees of freedom are 
identified and explicitly integrated out from the functional integral of the 
full theory. Using this method, the effective Lagrangian, which may be 
calculated to arbitrary subleading order in an inverse mass expansion, is 
derived for fields of spin-0, 1/2, and 1. 

\end{abstract}
\end{titlepage}

%================================ MAIN TEXT ===================================

\section{Introduction}

The application of quantum field theory to processes involving external 
states with a characteristic momentum scale $\Lambda$ is often complicated 
by the contributions from real and virtual excitations occuring at a much 
higher momentum of order $m \gg \Lambda$ which may preclude a straightforward 
perturbative treatment. The description of such phenomena with disparate 
scales can be greatly facilitated by the formulation of an effective field 
theory (EFT). In this approach, starting from a theory which describes 
processes at some high energy scale, heavy degrees of freedom, which are no 
longer dynamical at lower scales, are successively integrated out to yield 
an effective theory appropriate for the description of lower energy phenomena 
in terms of the remaining light excitations. 

A familiar effective field theory occurs in the description of  
electroweak phenomena. In the modern ${\rm SU}(2)_L \times {\rm U}(1)_Y$ 
theory which provides a description of the electroweak interactions up to 
energies of order hundreds of GeV, both matter fields and the weak 
gauge boson fields, $W^\pm$ and $Z^0$, appear explicitly. However, there 
are a multitude of weak processes which occur at energies of around a GeV, 
far below the gauge boson masses, where such heavy particles only appear 
as virtual degrees of freedom. For such low-momentum phenomena, it is 
convenient and advantageous to integrate out the $W^\pm$ and $Z^0$ fields 
to give an effective field theory in which the effects of these gauge 
particles appear as non-renormalizable fermionic operators suppressed by a 
mass of order the weak scale; the dominant subleading four-fermion operators 
reproduce the old Fermi theory of the weak interactions. 

In the following, we examine effective field theories of heavy particles 
which have some features that are qualitatively different from the example 
above. The aim of this presentation is to look at such theories in a more 
systematic and general context. For instance, the techniques to be illustrated 
here apply not just to elementary fields but also to those of composite 
particles as well as those of different spin. 

Let us begin by considering a quantum field theory which describes a 
particle with a mass $m$ which is coupled strongly to a non-Abelian gauge 
field at low energies (below $m$) so that it is not amenable to a 
perturbative treatment in this kinematic region. In addition, let $\LmdSI$ 
characterize the typical scale of the interactions.%
\footnote{A concrete example is the strong interactions of heavy quarks 
which will be discussed later in this paper \cite{PoW88,Gri90,Geo90,MRR91}.}

When the mass $m$ of this particle is taken to be very large compared to this 
scale, $m \gg \LmdSI$, with $\LmdSI$ fixed, there is a kinematic 
range below the large mass $m$ where the heavy particle is no 
longer fully dynamical --- for example, processes where heavy particle 
and antiparticle pair production occur are highly suppressed --- and 
particle and antiparticle number are separately 
conserved. Furthermore, when the magnitude of the typical momentum 
exchanged in such interactions and those involving other light 
degrees of freedom in the system is of order $\LmdSI$, the heavy particle 
propagates at some velocity $v$ (with $v^2 =1$), which is unaltered by 
these strong interactions, and its momentum can be expressed in the form
\eq p = m v + k, \endeql{ptot}
where $k$ is a ``residual momentum'' with magnitude of order 
$\LmdSI \ll m$ \cite{Geo90}. Interactions conserve the velocity and will 
only perturb $p$ by an amount of order $\LmdSI$. Typical momentum 
transfers are also of order $\LmdSI$. 

In this situation, it is convenient to integrate out the high-momentum 
excitations so that in the resulting effective field theory its 
effects are reproduced by higher-dimensional operators, which are 
accompanied by powers of $\LmdSI/m$, involving the remaining degrees 
of freedom. This theory would be valid for a single heavy particle in the 
kinematic region where low energy phenomena with momenta of order $\LmdSI$ 
take place. However, in this case, there is an important distinction from 
the above example of the electroweak theory: here, the heavy particle is 
real, and the heavy degrees of freedom 
which are removed correspond not to the entire heavy field but to the 
components of this field which decouple from the physical states in the 
infinite mass limit. Hence, external states in the EFT may still involve the 
heavy particle unlike the above case where they would only appear as 
virtual excitations. 

It is such an EFT, where the heavy particle may still be present as an 
external field, that we would like to construct starting from a full theory 
in a functional integral framework for particles of different spin. 
First, the heavy excitations will be identified and then explicitly 
integrated out from the path integral: since the particles and 
antiparticles decouple in the limit $m \to \infty$ independent of their 
spin, the heavy degrees of freedom correspond to either the particle or 
antiparticle components of the heavy field. In the subsequent 
analyses presented in this paper, the antiparticle component will be 
removed to give an EFT of heavy particles, but the procedure to obtain 
a heavy antiparticle EFT is virtually identical as we shall see. 

The result is an effective action which is usually non-local in the light 
fields that are left. Then expanding out this non-local action in an 
operator product expansion yields an infinite series of local operators of 
increasingly higher dimension suppressed by powers of the large scale. This 
procedure disentangles the low-energy physics, which may be non-perturbative, 
and is given by the structure of these operators from the effects of physics 
at high-energy which resides in the coefficients of these operators. In 
doing so, one also extracts the full dependence of physical observables on 
the heavy mass. In the following sections of this paper, such an approach 
will be used to formulate effective field theories for heavy particles of 
spin 0, 1/2, and 1. Although it may be intuitively evident that such 
effective theories should exist, it is nevertheless enlightening and 
reassuring to carry out this program explicitly.

\section{An Effective Field Theory of Heavy Scalar Particles}

Consider a non-Hermitian scalar field $\phi$ coupled strongly to a 
non-Abelian gauge field $\Acal^\mu$ described by the Lagrangian
\eq \Lagr_s = (D_\mu \phi)^\dagger D^\mu \phi - m_s^2\, \phi^\dagger \phi 
              +J^\dagger \phi +\phi^\dagger J,
\endeql{Lagr_s}
where
\eq D^\mu \phi = (\del^\mu - ig\Acal^\mu)\phi, \endeq
$J^\dagger$ and $J$ are external sources for $\phi$ and $\phi^\dagger$, 
respectively, and $m_s$ is the mass. \Eqr{Lagr_s} is the most general 
renormalizable Lagrangian in the absence of internal degrees of freedom 
(which may be incorporated straightforwardly) except for scalar 
self-interaction terms. Such terms are excluded because the object is to 
construct an EFT in the one-heavy-particle sector.
As in the above discussion, let $\LmdSI$ characterize the scale of the 
interactions, with $\LmdSI \ll m_s$. 
A concrete example of such a theory is one which is described by a chiral 
effective Lagrangian of a heavy pseudoscalar meson such as a $B$-meson 
interacting with a pseudo-Goldstone bosons where the interaction 
scale is $\Lambda_\chi \ll m_B$. However, to illustrate the methodology 
we shall continue to use the theory with the Lagrangian in \eqr{Lagr_s}. 

Since the conjugate momentum field of $\phi^\dagger$ is 
\eq \pi = {\del\Lagr_s \over \del{\dot\phi}^\dagger} = D^0 \phi, \endeq 
the Hamiltonian is given by 
\eq \Ham_s = \pi^\dagger {\dot\phi} + {\dot\phi}^\dagger \pi - \Lagr_s
  = \pi^\dagger \pi +(ig\Acal_0\,\phi)^\dagger \pi +\pi^\dagger ig\Acal_0\,\phi
      -(D_j \phi)^\dagger D^j \phi + m_s^2\, \phi^\dagger \phi
      -J^\dagger \phi -\phi^\dagger J. \endeq
The generating functional for the Green functions of this theory can be 
written as a functional integral over these fields:
\eqa Z[j_\phi,j_\phi^\dagger,j_\pi,j_\pi^\dagger] \hskip -.5em
 &=& \hskip -.5em
   N\int e^{i\int (\pi^\dagger {\dot\phi} + {\dot\phi}^\dagger \pi -\Ham_s 
   +j_\phi^\dagger \phi +\phi^\dagger j_\phi +j_\pi^\dagger \pi +\pi^\dagger 
    j_\pi) d^4x}\, \D\phi\, \D\phi^\dagger\, \D\pi\, \D\pi^\dagger\,\nonumber\\
 &=& \hskip -.5em
   N\int e^{i\int [\pi^\dagger (D_0\phi) +(D_0\phi)^\dagger \pi -\pi^\dagger
      \pi +(D_j\phi)^\dagger (D^j\phi) -m_s^2 \phi^\dagger \phi 
     +(j_\phi +J)^\dagger \phi +\phi^\dagger (j_\phi +J) +j_\pi^\dagger \pi 
     +\pi^\dagger j_\pi] d^4x} \nonumber \\
 &&\hskip 1.5em \D\phi\, \D\phi^\dagger\, \D\pi\, \D\pi^\dagger\, \endeqa
where $N$ is a normalization constant. 

This procedure may be generalized to a frame moving with velocity $v^\mu$ 
where the conjugate momentum field of $\phi^\dagger$ is now
\eq \pi_v = {\del\Lagr_s \over \del(v\cdot \del\phi)^\dagger} = \vD \phi.
\endeq
Then the generating functional becomes 
\eqa Z[j_\phi,j_\phi^\dagger,j_{\pi_v},j_{\pi_v}^\dagger] 
 &=&N\int e^{i\int [\pi_v^\dagger (v\cdot\del\phi) + (v\cdot\del\phi)^\dagger 
     \pi_v -\Ham_s
     +j_\phi^\dagger \phi +\phi^\dagger j_\phi +j_{\pi_v}^\dagger \pi_v
     +\pi_v^\dagger j_{\pi_v}] d^4x} \,
             \D\phi\, \D\phi^\dagger\, \D\pi_v\, \D\pi_v^\dagger\, \nonumber\\
 &=&N\int e^{i\int [L_s(\phi,\pi) +j_\phi^\dagger \phi +\phi^\dagger j_\phi 
            +j_{\pi_v}^\dagger \pi_v +\pi_v^\dagger j_{\pi_v}] d^4x}\,
              \D\phi\, \D\phi^\dagger\, \D\pi_v\, \D\pi_v^\dagger, \endeqa
where 
\eq L_s(\phi) = \pi_v^\dagger (\vD\phi) +(\vD\phi)^\dagger \pi_v 
   -\pi_v^\dagger \pi_v +(\Dperp_\mu\phi)^\dagger (D^{\perp\mu}\phi) 
   -m_s^2\, \phi^\dagger \phi +J^\dagger \phi +\phi^\dagger J, \endeq
and $\Dperp_\mu$ is the component of the covariant derivative orthogonal 
to the direction of the velocity $v^\mu$:
\eq \Dperp_\mu = D_\mu - v_\mu (\vD) \endeq

To identify the heavy degrees of freedom, first observe that in the large 
mass limit, the (predominantly) heavy particle field (with positive energy) 
is given by the projection
\alpheq
\eq \phi^+ = {1\over 2} \left(1 +{i \vD \over m_s}\right)\phi,\endeq
while the (predominantly) heavy antiparticle field (with negative energy) 
is given by 
\eq \phi^- = {1\over 2} \left(1 -{i \vD \over m_s}\right)\phi,\endeq
\reseteq
so that 
\alpheq
\eq \phi = \phi^+ + \phi^-,\endeq
\eq {\left(i \vD \over m_s\right)}\phi = \phi^+ - \phi^-.\endeq
\reseteq

Making the decomposition into particle and antiparticle components 
yields 
\eq Z[j_{\phi^+},j_{\phi^+}^\dagger,j_{\phi^-},j_{\phi^-}^\dagger] 
   = N \int e^{i\int [L_s(\phi^+, \phi^-) +(j_{\phi^+}^\dagger \phi^+ 
        +j_{\phi^-}^\dagger \phi^- + {\rm h.c.})]d^4 x}\,
      \D\phi^+\, \D\phi^-\, \D(\phi^+)^\dagger\, \D(\phi^-)^\dagger,\endeq
where
\eqa L_s(\phi^+, \phi^-) 
 &=& 2m_s i[(\phi^+)^\dagger \vD \phi^+ -(\phi^-)^\dagger \vD \phi^-]
   -2m_s^2 [(\phi^+)^\dagger \phi^+ +(\phi^-)^\dagger \phi^-] \nonumber \\
 &&+[\Dperp_\mu (\phi^+ +\phi^-)]^\dagger D^{\perp\mu} (\phi^+ +\phi^-)
   +J^\dagger (\phi^+ +\phi^-) +(\phi^+ +\phi^-)^\dagger J,
\endeqa
and quantities independent of the fields have been absorbed into $N$. 
Note that this function $L_s$ which appears in the generating functional 
is {\it different}\/ from the original Lagrangian $\Lagr_s$.
The ``h.c.'' denotes hermitian conjugate terms. 

To arrive at an EFT of heavy scalars, the antiscalar component must be 
integrated out. However, it is useful to first remove from the total 
momentum of the heavy field the large momentum piece $m_s v$ in 
\eqr{ptot} by defining a new field $\phi_v$ at a velocity $v$:
\alpheq
\eq \phi(x) = e^{-i m_s v\cdot x} \phi_v(x) \endeql{phiv}
and similarly for the component fields
\eq \phi^\pm(x) = e^{-i m_s v\cdot x} \phi_v^\pm(x).\endeql{phiv+-}
\reseteq
Now, derivatives acting on $\phi_v$ only give factors of the residual 
momentum $k$ 
and thus facilitating a systematic derivative expansion of operators in 
powers of $k/m_s \sim \LmdSI /m_s$. To arrive at an EFT of heavy 
antiparticles, the factor $e^{-i m_s v\cdot x}$ would be replaced by 
$e^{+i m_s v\cdot x}$ in eq.~(\ref{phiv}--\ref{phiv+-}); for particles 
with spin this is also the appropriate replacement 
($v \rightarrow -v$) together with a suitable change in the mass. 
Implementing these transformations in the above generating functional 
gives 
\eqa Z[j_{\phi^+},j_{\phi^+}^\dagger,j_{\phi^-},j_{\phi^-}^\dagger] 
  &=& N \int e^{i\int [L_s^v(\phi_v^+, \phi_v^-) 
      +(j_{\phi^+}^\dagger e^{-im_s v\cdot x}\phi_v^+ 
        +j_{\phi^-}^\dagger e^{-im_s v\cdot x}\phi_v^- + \hc)]d^4 x}\nonumber\\
  &&\hskip 2em
     \D\phi_v^+\, \D\phi_v^-\, \D(\phi_v^+)^\dagger\, \D(\phi_v^-)^\dagger,
\endeqa
where
\eqa L_s^v(\phi_v^+, \phi_v^-) 
 &=& 2m_s [(\phi_v^+)^\dagger i\vD \phi_v^+ 
        -(\phi_v^-)^\dagger (2m_s +i\vD) \phi_v^-]
     -(\phi_v^+ +\phi_v^-)^\dagger (\Dperp)^2 (\phi_v^+ +\phi_v^-) \nonumber\\
  && +J^\dagger e^{-im_s v\cdot x} (\phi_v^+ +\phi_v^-)
     +e^{im_s v\cdot x} (\phi_v^+ +\phi_v^-)^\dagger J, \endeqal{L_s^v}
Now setting the sources for the $\phi_v^-$ and the $\phi_v^{-\dagger}$ 
fields to zero, $j_{\phi^-} =j_{\phi^-}^\dagger =0$, and performing 
the functional integral over these fields yields the result
\eq Z[j_{\vphiv},j_{\vphiv}^\dagger,0,0] 
   = N \hskip -1mm \int \hskip -1mm e^{i\int [L_s^{'v}(\vphiv)
         +(j_{\vphiv}^\dagger \vphiv + \hc)]d^4 x}
     \{\det i[2m_s (2m_s +i\vD) +(\Dperp)^2]\}^{-1} 
     \D\vphiv \D\vphiv^\dagger, \endeql{ZSFTnonlocv}
where one has used the simplified notation
\eq \vphiv = \phi_v^+.\endeq
In \eqr{ZSFTnonlocv} 
\eqa L_s^{'v}(\vphiv)
   &=& \vphiv^\dagger [2m_s\: i\vD -(\Dperp)^2]\vphiv
       +J^\dagger e^{-im_s v\cdot x} \vphiv
       +\vphiv^\dagger e^{im_s v\cdot x} J \nonumber \\
   && +[-\vphiv^\dagger (\Dperp)^2 +J^\dagger e^{-im_s v\cdot x}]
      [2m_s (2m_s +i\vD) +(\Dperp)^2]^{-1} \nonumber \\
   &&\hskip 1em [-(\Dperp)^2 \vphiv +J e^{im_s v\cdot x}],\endeqal{LHSEFTv}
and
\eq j_{\vphiv} = j_{\phi_v^+} = e^{im_s v\cdot x} j_{\phi^+} \endeq
is the source for $\vphiv^\dagger$. The determinant factor in \eqr{ZHSEFTv} 
is a consequence of quantum effects. 

\Eqr{LHSEFTv} clearly contains non-local terms, but now one may 
systematically expand in powers of derivatives over the large mass to 
arrive at the heavy scalar effective field theory (HSEFT) Lagrangian
\eqa \Lagr_{\rm HSEFT}^v(\vphiv)
 &=& \vphiv^\dagger [2m_s\: i\vD -(\Dperp)^2)]\vphiv
     +J^\dagger e^{-im_s v\cdot x} \vphiv
     +\vphiv^\dagger e^{im_s v\cdot x} J \nonumber \\
 &&  +{1\over 4m_s^2}[\vphiv^\dagger (\Dperp)^4 \vphiv
       -\vphiv^\dagger (\Dperp)^2 J e^{im_s v\cdot x}
       -J^\dagger e^{-im_s v\cdot x} (\Dperp)^2 \vphiv +J^\dagger J]\nonumber\\
 &&-{1\over 8m_s^3}
    \Biggl\{ [-\vphiv^\dagger (\Dperp)^2 +J^\dagger e^{-im_s v\cdot x}]
     \Biggl[i\vD +{(\Dperp)^2 \over 2m_s}\Biggr] 
     [-(\Dperp)^2 \vphiv +J e^{im_s v\cdot x}] \Biggr\} \nonumber \\
 &&  +{\cal O}\left({1\over m_s^4}\right), \endeqal{LagrHSEFTv}
with the generating functional
\eqa Z_{\rm HSEFT}[j_{\vphiv},j_{\vphiv}^\dagger,0,0] 
   &=& N \hskip -1mm \int \hskip -1mm e^{i\int [\Lagr_{\rm HSEFT}^v(\vphiv)
         +(j_{\vphiv}^\dagger \vphiv + \hc)]d^4 x} \nonumber\\
     &&\hskip 2em \{\det i[2m_s (2m_s +i\vD) +(\Dperp)^2]\}^{-1} 
       \, \D\vphiv \D\vphiv^\dagger. \endeqal{ZHSEFTv}
When the theory was expressed in terms of velocity-dependent fields above, 
a {\it particular}\/ velocity $v$ was selected which breaks the Lorentz 
covariance of the theory. Furthermore, since the different velocity 
sectors are not coupled to one another by the ``velocity superselection 
rule'' \cite{Geo90}, in order to recover Lorentz covariance all possible 
velocities should be included so that the complete generating functional 
becomes
\eqa \Z[j_{\vphiv},j_{\vphiv}^\dagger] 
   &=& N \int e^{i\int \sum_v[\Lagr_{\rm HSEFT}^v(\vphiv)
         +(j_{\vphiv}^\dagger \vphiv + \hc)]d^4 x} \nonumber\\
     &&\hskip 2em \prod_v \{\det i[2m_s (2m_s +i\vD) +(\Dperp)^2]\}^{-1}
        \, \D\vphiv \D\vphiv^\dagger. \endeqal{ZHSEFT}

In this HSEFT as it has been formulated here, $\vphiv$ only acts 
on scalars and not on antiscalars. For a theory with antiscalars, they 
would have to be included separately through the transformation indicated 
above. Moreover, additional flavours of heavy scalars (each with mass 
$m_{s_i} \gg \LmdSI$) are readily incorporated into the above formalism: 
the complete generating functional is then the product of the generating 
functionals for each species $i$ and the corresponding effective Lagrangian 
is the sum of the individual ones. Hence if the fields were scaled as
\eq \vphiv = {\vphiv' \over \sqrt{2m_s}}, \endeq
then for $N_s$ heavy scalar flavours the leading order effective Lagrangian 
in \eqr{LagrHSEFTv} would have a SU($N_s$) symmetry. Operator insertions 
can also be readily accommodated by adding to the full theory Lagrangian, 
\eqr{Lagr_s}, a term with the operator coupled to a source. 

The calculation performed here yields the tree-level effective Lagrangian, 
\eqr{LagrHSEFTv}. This quantity can also be derived by using the classical 
equation of motion for $\phi_v^-$ from \eqr{L_s^v}, namely 
\eq \phi_v^- = -\left[2m_s (2m_s +i\vD) +(\Dperp)^2 \right]^{-1}
 \left[(\Dperp)^2 \phi_v^+ -e^{im_s v\cdot x} J \right], \endeql{phiEoM}
to express $\phi_v^-$ in terms of $\phi_v^+$ in that Lagrangian. However, 
these two approaches will differ when quantum effects are included and 
herein lies an advantage of the functional integral approach where such 
contributions can be incorporated methodically. In particular, the relation 
between $\phi_v^-$ and $\phi_v^+$ in \eqr{phiEoM} will be altered by such 
effects. The equation of motion method also fails to generate the 
determinantal factor in \eqr{ZHSEFTv}. To calculate physical quantities when 
radiative corrections are taken into account, it is necessary to choose a 
suitable regularization and renormalization scheme. However, since the 
choice of such schemes is the same for the theories considered in this paper, 
a discussion of this subject will be prosponed until the following section 
where we examine heavy spin-$\smlfrac{1}{2}$ particles because they occur 
in some theories of considerable interest and thus affords us the 
opportunity to treat them in a physically realized setting. 

Finally, since any observable may be expressed in terms of a Green function 
which are, in turn, all generated by the action functional, \eqr{ZHSEFTv} 
or (\ref{ZHSEFT}), these equations along with a regularization and 
renormalization scheme provide a complete framework for performing 
calculations in this theory.

\section{An Effective Field Theory of Heavy Spin-${1 \over 2}$ Fermions}

There are a number of examples in nature of spin-$\smlfrac{1}{2}$ fermions 
whose masses are large compared with their characteristic interaction 
energies: for instance, heavy $b$ and $c$ quarks in QCD, and the chiral 
interactions of heavy spin-$\smlfrac{1}{2}$ baryons with pseudo-Goldstone 
bosons amongst others. In this paper, we will apply the above method to the 
low-momentum interactions of heavy quarks in QCD \cite{MRR91};
applications of this methodology to the chiral interactions of heavy baryons 
will be presented in a subsequent publication.

The strong interactions of a given flavour of heavy quark, having a 
mass $m_Q \gg \LmdQCD$, with coloured gluons $\Acal^\mu$ and coupled to 
an external source $\zeta$ is described by the Lagrangian 
\eq \Lagr_{\rm HQ} = \bar \psi (i\Dslash -m_Q) \psi +\bzeta \psi 
 +\bar\psi \zeta. \endeql{L_H-QCD}
$\psi$ is the heavy quark field in QCD and $D^\mu$ is the gauge-covariant 
derivative:
\eq D^\mu \psi= (\partial^\mu - ig\Acal^\mu_a T^a) \psi \endeq
The gluon field tensor $G^{\mu \nu}_a$ is defined by
\eq [D^\mu,D^\nu] = -igG^{\mu \nu}_a T^a, \endeq
where $T^a$ is the colour SU(3) generator.

The action functional in a frame moving at velocity $v^\mu$ with sources 
$\eta$ and $\bar\eta$ for $\bpsi$ and $\psi$, respectively, is
\eq Z[\eta,\bareta]
 = N\int e^{i\int [\rho^\dagger v\cdot \del\psi + v\cdot \del\psi^\dagger \rho
     -\Ham_{\rm HQ} +\bareta \psi +\bpsi \eta]d^4 x}\, \D\psi \D\bpsi,
\endeql{ZQ1}
where
$\rho^\dagger$ is the conjugate momentum field of $\psi$,
\eq \rho^\dagger = {\del\Lagr_{\rm HQ} \over \del(v\cdot\del\psi)}
 = \smlfrac{i}{2} \bpsi \vslash,\endeq
and $\Ham_{\rm HQ}$ is the Hamiltonian:
\eq \Ham_{\rm HQ} = -i \bpsi \Dperpsl \psi +m_Q\bpsi \psi
                    -\bzeta \psi -\bpsi \zeta \endeq
\Eqr{ZQ1} can be simplified to read
\eq Z[\eta,\bareta]
 = N\int e^{i\int [L_{\rm HQ}^v +\bareta \psi +\bpsi \eta]d^4 x}\,
   \D\psi\, \D\bpsi,
\endeql{ZQ2}
where
\eq \LvHQ = 
 \bpsi [i(\vslash \vD + \Dperpsl) -m_Q]\psi
 +\bzeta \psi +\bpsi \zeta. \endeql{LvHQ}
Although in this case the quantity appearing in the generating functional of 
\eqr{ZQ2} is $\LvHQ$ which coincides with the original Lagrangian 
$\Lagr_{\rm HQ}$, this is generally not true whenever there 
are two or more time derivatives in kinetic terms of the Lagrangian as 
one may see from the above analysis of the scalar field theory. 

As in the usual treatment of heavy quarks in an effective field theory, 
the heavy quark and heavy antiquark components at a velocity 
$v^\mu$, $\psi_v^+$ and $\psi_v^-$, respectively, are defined by 
\eq \psi_v^\pm = P_v^\pm \psi, \endeq
with 
\eq P_v^\pm = {1 \pm \vslash \over 2}, \endeq
\eq \psi = \psi_v^+ + \psi_v^-.\endeq
Then to obtain an effective theory for heavy quarks, the kinematic 
dependence of the fields on the heavy mass is removed by the 
transformation 
\eq \psi_v^\pm (x) = e^{-im_Q v\cdot x} h_v^\pm (x),\endeq
with
\eq h_v = \hvpos + \hvneg.\endeq
Implementing these changes gives
\eqa Z[\eta,\bareta] \hskip -.5em
 &=&\hskip -.5em
   N\int e^{i\int [\LvHQ(\psi,\bpsi) +\bareta \psi +\bpsi \eta]d^4 x}\,
   \delta(\hvpos -e^{im_Q v\cdot x} P_v^+ \psi) \,
   \delta(\hvneg -e^{im_Q v\cdot x} P_v^- \psi) \nonumber \\
 &&\hskip 1em \delta(\bar\hvpos -e^{-im_Q v\cdot x} \bpsi P_v^+) \,
   \delta(\bar\hvneg -e^{-im_Q v\cdot x} \bpsi P_v^-) \,
   \D\psi\,\D\bpsi\,\D\hvpos\, \D\hvneg\, \D\bar\hvpos\, \D\bar\hvneg.\endeqa
Defining
\eq \eta_v^\pm = e^{im_Q v\cdot x} P_v^\pm \eta,
 \qquad \zeta_v^\pm = e^{im_Q v\cdot x} P_v^\pm \zeta, \endeq
for the sources for $h_v^\pm$, and integrating over the $\psi$ and $\bpsi$ 
fields gives the action functional in terms of the new $h_v^\pm$ fields:
\eq Z[\eta_v^+,\eta_v^-,\bar\eta_v^+,\bar\eta_v^-]
 = N\int e^{i\int [\LvHQ(h_v^\pm,\bar h_v^\pm) 
    +L_{\rm source}(h_v^\pm,\bar h_v^\pm;\eta_v^\pm,\bar\eta_v^\pm)]d^4 x}\,
   \D\hvpos\, \D\hvneg\, \D\bar\hvpos \D\bar\hvneg, \endeq
where
\eqa
 \LvHQ(h_v^\pm,\bar h_v^\pm)
 &=& \bhvpos i\vD \hvpos -\bhvneg (i\vD +2m_Q)\hvneg
   +\bhvpos i\Dperpsl \hvneg +\bhvneg i\Dperpsl \hvpos \nonumber\\
 &&  +(\bzeta_v^+ \hvpos +\bzeta_v^- \hvneg +{\rm h.c.}), \\
 L_{\rm source} &=& \bar\eta_v^+ h_v^+ +\bar\eta_v^- h_v^- 
   +\bar h_v^+ \eta_v^+ +\bar h_v^- \eta_v^-. \endeqa
As in the scalar case, the heavy antiquark component is the heavy degree of 
freedom to be removed, so setting the 
corresponding sources to zero, $\eta_v^- = \bar\eta_v^- = 0$, 
and integrating over $h_v^-$ and $\bar h_v^-$ yields the generating 
functional 
\eq Z[\eta_v^+,\bar\eta_v^+]
 = N\int e^{i\int [L^v_{\rm HQEFT}(\hvpos,\bhvpos) 
    +L_{\rm source}(\hvpos,\bhvpos;\eta_v^+,\bar\eta_v^+)]d^4 x} \,
   \det(2m_Q +i\vD)\, \D\hvpos\, \D\bhvpos,\endeql{ZHQEFT}
where
\eq L^v_{\rm HQEFT} = \bhvpos i\vD \hvpos 
 +(\bhvpos i\Dperpsl +\bzeta_v^-)(2m_Q +i\vD)^{-1}
  (i\Dperpsl \hvpos +\zeta_v^-) +(\bzeta_v^+ \hvpos +{\rm h.c.}),\endeq
and
\eq L_{\rm source}(\hvpos,\bhvpos;\eta_v^+,\bar\eta_v^+) 
 = \bar\eta_v^+ \hvpos +\bhvpos \eta_v^+.\endeq
The determinant in \eqr{ZHQEFT} arises from integrating out the quantum 
fluctuations of the $h_v^-$ field. This quantity may be regulated so 
that gauge invariance is preserved, and when it is evaluated in an axial 
gauge with $v\cdot \Acal = 0$, it turns out to be constant \cite{MRR91}. 

As before, integrating out the heavy degrees of freedom leads to a 
non-local effective Lagrangian, but one which has a systematic 
derivative expansion in powers of $\LmdQCD/m_Q$ and where short and 
long distance scales are separated; this heavy quark effective field 
theory (HQEFT) Lagrangian is
\eq \LHQEFT^v = \sum_{n=0}^\infty \LHQEFT^{v(n)} , \endeql{L_HQEFT,v}
where the superscript $n$ denotes the $n$th order term in the
$1/m_Q$ expansion of $\LHQEFT^v$. The first several terms (with the sources 
set to zero) are 
\eqa
 \LHQEFT^{v(0)} &=& \bar Q_v i(\vD) Q_v , \label{L^(0)}\nonumber\\
 \LHQEFT^{v(1)} &=& \frac{1}{2m_Q} \bar Q_v 
   \left[(iD)^2 +\frac{g}{2} \sigma^{\mu \nu} G_{\mu\nu} -(i\vD)^2 \right] Q_v,
  \label{L^(1)}\nonumber\\
 \LHQEFT^{v(2)} &=& \frac{i}{4m_Q^2} 
   \bar Q_v [\Dslash (\vD) \Dslash -(\vD)^3] Q_v, \nonumber\\
  &=& \frac{1}{4m_Q^2} \bar Q_v
   \biggl[{1\over 2}g v^\mu [D^\nu, G_{\mu \nu}] 
   +{ig\over 2}\sigma^{\alpha \mu} v^\nu \{D_\alpha, G_{\mu\nu}\} \cr
   && \hskip 5em +{i\over 2}\bigl\{D^2 -{g\over 2}\sigma^{\mu\nu} G_{\mu\nu}, 
      \vD\bigr\} -(\vD)^3 \biggr] Q_v, \label{L^(2)}\nonumber\\
 \LHQEFT^{v(3)} &=& \frac{1}{8m_Q^3} \bar Q_v
  [\Dslash (\vD)^2\, \Dslash - (\vD)^4] Q_v. \label{L^(3)}
\endeqal{L_HQEFT,vi}
And the HQEFT generating functional is given by \eqr{ZHQEFT} with 
$L^v_{\rm HQEFT}$ replaced by $\LHQEFT^v$. 

Just as for scalar fields, Lorentz covariance of this effective theory is 
recovered by analogously including the various velocity sectors leading 
to the effective Lagragian 
\eq \LHQEFT = \sum_v \LHQEFT^v, \endeql{L_HQEFT}
which at leading-order has the well-known SU($2N_f$) spin-flavour symmetry 
for $N_f$ flavours of heavy quarks \cite{IsW89&90}.

The path integration over the heavy excitations gives the HQEFT Lagrangian, 
\eqr{L_HQEFT,v} and (\ref{L_HQEFT}), which reproduces QCD at tree-level for 
scales below $m_Q$. In doing so, all of the internal heavy quark loops have 
been integrated out from the theory. However, as we had alluded to in the 
previous section, to completely specify an EFT requires, in addition, the 
specification of a regularization and renormalization scheme. Since 
dimensional regularization preserves all of the physical properties of the 
theory except that space-time is no longer four-dimensional, it is the most 
suitable choice for massive particles coupled to gauge fields. 
Perhaps the most convenient renormalization scheme is one involving a 
mass-independent subtraction (such as MS or $\rm{\overline{MS}}$) and it 
is the one employed here.%
\footnote{Although in a mass-independent subtraction scheme the heavy 
particles do not decouple \cite{ApC75}, this does not present a problem here 
because below its mass $m_Q$ the dynamical degrees of freedom of the heavy 
quark have been explicitly integrated out.}

Since the high energy behaviour of HQEFT is different from that of QCD, 
when radiative contributions are included, the HQEFT Lagrangian must be 
corrected from its tree-level form by introducing short distance 
coefficients for the operators in \eqr{L_HQEFT,vi} which are determined 
by matching physical quantities calculated in the two theories. 
Since the matching is generally performed at the heavy mass thresholds, 
the values of the coefficients at lower scales are determined by solving 
for their evolution as governed by the renormalization group equations.%
\footnote{See for instance ref. \cite{PoW88}.}

\section{An Effective Field Theory of Heavy Vector Particles}

In nature, there are some instances where the low-momentum behaviour 
of heavy spin-1 particles may be best described by an effective field 
theory such as the chiral interactions of $D^*$ or $B^*$ vector 
mesons with pseudo-Goldstone bosons. For simplicity, we shall illustrate 
the formulation of an effective field theory for a heavy vector field 
$A^\mu$ with mass $m_V$ described by the Lagrangian 
\eq \Lagr_V = -\frac{1}{2} 
               (D_\mu A_\nu -D_\nu A_\mu)^\dagger (D^\mu A^\nu -D^\nu A^\mu) 
              +(m_V)^2 A_\mu^\dagger A^\mu,\endeql{L_V}
where the covariant derivative prescribes the interaction of the massive 
vector with the gauge field
\eq D^\mu A^\nu = (\del^\mu - ig\Acal^\mu)A^\nu, \endeq
with a typical interaction scale of $\LmdSI$.
The equation of motion for the field is 
\eq D_\mu (D^\mu A^\nu -D^\nu A^\mu) +(m_V)^2 A^\nu = 0. \endeql{constraint}
Without internal symmetries (which may be included), \eqr{L_V} is the most 
general Lagrangian when self-interaction terms for the heavy field are 
excluded (as they are irrelevant in the one-heavy-particle sector). 
The procedure given below can be used to derive a heavy particle effective 
field theory for more complicated and physically realized cases such as 
the example given above.%
\footnote{Some instances where such a theory is used in chiral interactions 
are ref. \cite{JMW95,BGT97}.}

We shall first obtain the Hamiltonian which will be needed subsequently. 
In a coordinate frame with velocity $v^\mu$, the conjugate momentum field to 
$A_\nu^\dagger$ is
\eq \Pi^\nu = {\del\Lagr_V \over \del(v\cdot\del A_\nu^\dagger)}
 = -v_\mu \Gcal^{\mu\nu} 
 = -v_\mu (\Gcal^{\parallel\mu\nu} +\Gcal^{\perp\mu\nu}),\endeq
where
\eq \Gcal^\parallel_{\mu\nu}
 = (v_\mu \, \vD)A_\nu -(v_\nu \, \vD)A_\mu,\endeq
and
\eq \Gcalperp_{\mu\nu} = \Dperp_\mu A_\nu -\Dperp_\nu A_\mu, \endeq
so the Hamiltonian may then be written as 
\eqa \Ham_V 
 &=&-\Pi_\nu^\dagger \Pi^\nu -[\Pi^{\nu\dagger} v^\mu \Gcalperp_{\mu\nu}
    +\Pi^{\nu\dagger} v\cdot(-ig\Acal)A_\nu +\hc]
    -v^\mu \Gcalperp_{\mu\nu}^\dagger v_\alpha \Gcalperp^{\alpha\nu}\nonumber\\
 && +\frac{1}{2}\Gcalperp_{\mu\nu}^\dagger \Gcalperp^{\mu\nu}
    -(m_V)^2 A_\mu^\dagger A^\mu. \endeqa

This theory is singular as one may see from the fact that in the rest frame 
the $A^0$ component has no conjugate momentum field, and consequently, there 
are constraints. The Hamiltonian formalism, in which such constraints are 
taken into account, will be employed to quantize the system 
\cite{quantconstrnt}. In this moving frame there is a primary constraint:
\eq \Phi^{(1)} = v\cdot \Pi = 0 \endeq
Secondary constraints may be derived by requiring that this primary 
constraint is consistent with the equations of motion:
\eq v\cdot \del\, \Phi^{(1)} = \left\{\Phi^{(1)},\Ham_V \right\} = 0 \endeq
Commuting the primary constraint with the Hamiltonian gives
\eq \Phi^{(2)} =-\vD (v\cdot \Pi) +\Dperp_\mu \Pi^\mu +(m_V)^2\, v\cdot A.
\endeql{constraint2}
Since massive spin-1 particles have only three physical degrees of freedom 
and it is represented here as a four-component vector field, 
\eqr{constraint2} serves to eliminate the spurious degree of freedom.
There are no other secondary constraints, and $\Phi = (\Phi^{(1)},\Phi^{(2)})$
is the full system of constraints. The matrix consisting of the Poisson 
bracket of all constraints, namely
\eqa \{\Phi({\bf x}),\Phi({\bf y})\} = 
 \left( \begin{array}{cc}
         0 & (m_V)^2 \\
         -(m_V)^2 & 0 
        \end{array} \right)
 \delta ({\bf x} - {\bf y}), \endeqal{constrtmatr}
is nonsingular here, so this is a theory with second-class constraints.

The generating functional for the heavy vector field $A^\mu$ and the 
conjugate momentum field $\Pi^{\nu\dagger}$ with the corresponding sources 
$j_A^\mu$ and $j_\Pi^\mu$ which implements these constraints is
\eqa Z[j_A^\mu,j_A^{\mu\dagger},j_\Pi^\mu,j_\Pi^{\mu\dagger}]
 &=& N\int e^{i\int [L_V (A_\mu,\Pi_\mu;\hc) 
       +(j_A^{\mu\dagger} A_\mu +j_\Pi^{\mu\dagger} \Pi_\mu + \hc)] d^4x}
      \nonumber\\
 &&\hskip 2em \det^\smlfrac{1}{2} \{\Phi,\Phi\} \>
      \delta(v\cdot\Pi)\, \delta(v\cdot\Pi^\dagger) \nonumber\\ 
 &&\hskip 2em \delta(-\vD(v\cdot\Pi) +\Dperp_\mu \Pi^\mu
       +(m_V)^2 v\cdot A) \nonumber \\
 &&\hskip 2em \delta[(-\vD(v\cdot\Pi) +\Dperp_\mu \Pi^\mu +(m_V)^2%
        v\cdot A)^\dagger]\nonumber \\
 &&\hskip 2em \D A_\mu\, \D A_\mu^\dagger\, \D\Pi_\mu\, \D\Pi_\mu^\dagger, 
\endeqa
where 
\eq L_V (A^\nu,\Pi^\nu) = \Pi^{\nu\dagger} (v\cdot\del A_\nu)
              +(v\cdot\del A_\nu)^\dagger \Pi^\nu -\Ham_V(A^\nu,\Pi^\nu).\endeq
Because the matrix $\{\Phi,\Phi\}$ in \eqr{constrtmatr} is independent of 
the fields, the corresponding determinant factor 
$\det^\smlfrac{1}{2} \{\Phi,\Phi\}$ can be factored out of the path 
integral and henceforth will be absorbed into the normalization. 

It is convenient to change variables into quantities defined with respect to 
the velocity by introducing the following projectors parallel and 
perpendicular projectors
\alpheq
\eqa \Ppara = v_\mu v_\nu \\
     \Pperp = g_{\mu\nu} -v_\mu v_\nu. \endeqa
\reseteq
Hence, one defines the component of the vector field parallel 
to the velocity as
\alpheq
\eq \Apara = v\cdot A, \endeq
and the perpendicular component to be
\eq \Aperp_\mu = \Pperp A_\nu = A_\mu - v_\mu \Apara, \endeql{Aperpdef}
with the constraint
\eq v\cdot \Aperp = 0, \endeq
so that 
\eq A_\mu = (\Ppara +\Pperp)A_\nu = v_\mu\Apara + \Aperp_\mu. \endeq
\reseteq
Similarly, the parallel and perpendicular components of the conjugate 
momentum field are defined to be, respectively, 
\alpheq
\eqa \Pipara = v\cdot \Pi,\\
     \Piperp_\mu = \Pperp \Pi_\nu = \Pi_\mu - v_\mu \Pipara, \endeqa
with
\eqa v\cdot \Piperp = 0,\\
     \Pi_\mu = (\Ppara +\Pperp)\Pi_\nu = v_\mu\Pipara + \Piperp_\mu. \endeqa
\reseteq
Making this change of variables and then integrating over the fields 
$\Pipara, {\Pipara}^\dagger$ and implementing the above delta function 
constraints in $L_V$ yields
\eqa
 Z[j_A^{\mu\perp},j_A^{\mu\perp\dagger},j_\Pi^{\perp\mu},
   j_\Pi^{\perp\mu\dagger}]
 &=&N\int e^{i\int [L_V (\Aperp_\mu,\Apara,\Piperp_\nu;\hc)
    +(j_A^{\mu\perp\dagger}\Aperp_\mu +j_\Pi^{\perp\mu\dagger}\Piperp_\mu
      +\hc)]d^4x} \nonumber \\
 &&\hskip 2em [\delta(v\cdot\Aperp)\, \D\Aperp_\mu\, \D\Apara\,
       \delta(v\cdot\Piperp)\, \D\Piperp_\mu \times (\hc)], \endeqa
where
\eqa L_V (\Aperp_\mu,\Apara,\Piperp_\nu;\hc)
 &=& (\Pi^{\mu\perp})^\dagger (\vD\Aperp_\mu -\Dperp_\mu \Apara)
     +(\vD\Aperp_\mu -\Dperp_\mu \Apara)^\dagger \Pi^{\mu\perp} \nonumber \\
 & & +(\Piperp_\mu)^\dagger \Pi^{\mu\perp}
     +\frac{1}{2} (\Dperp_\mu \Aperp_\nu -\Dperp_\nu \Aperp_\mu)^\dagger
      (D^{\perp\mu} A^{\nu\perp} -D^{\perp\nu} A^{\mu\perp}) \nonumber \\
 & & +(m_V)^2 [(\Aperp_\mu)^\dagger A^{\mu\perp} +(\Apara)^\dagger \Apara].
\endeqa
% Insert discussion of y Z now fn of perp quants only.

The next step is to identify the heavy degrees of freedom. In the heavy 
mass limit, the heavy vector $A_\mu^+$ and heavy antivector $A_\mu^-$ 
component fields may be identified with 
\eq A_\mu^\pm(x)
 = \frac{1}{2} \left(1 \pm \frac{i\vD}{m_V}\right) A_\mu(x),
\endeq
The original field $A^\mu$ and its derivative may then be expressed in terms 
of these components through
\alpheq
\eq A_\mu = A_\mu^+ + A_\mu^-, \endeq
\eq \left(\frac{i\vD}{m_V}\right) A_\mu = A_\mu^+ - A_\mu^-. \endeq
\reseteq
These equations can then be used to reexpress the action functional in 
terms of the positive and negative energy components $A_\mu^\pm$. 

It is clear that to obtain an EFT which describes heavy vectors, $A_\mu^-$ 
should be integrated out. As before, however, it is convenient to first 
remove the kinematic dependence of the heavy field on $m_V$ by defining 
new velocity-dependent fields:
\eqa A_\mu(x) = e^{-i m_V v\cdot x} A_{\mu,v}(x), \nonumber\\
     A_\mu^{\perp\pm}(x) = e^{-i m_V v\cdot x} A_{\mu,v}^{\perp\pm}(x),\\
     \Apara(x) = e^{-i m_V v\cdot x} \Apara_v(x). \nonumber \endeqa
Expressing the generating functional in terms of these new quantities and 
integrating out $\Apara,A_{\mu,v}^{\perp -}$ as well as their hermitian 
conjugate fields yields 
\eqa Z[j_{A,v}^{\mu\perp +},(j_{A,v}^{\mu\perp +})^\dagger]
 &=&N\int e^{i\int [L_{\rm HVEFT}(\Avperpl{\mu},(\Avperpl{\mu})^\dagger)
     +(j_{A,v}^{\mu\perp +} A_{\mu,v}^{\perp +} +\hc)] d^4x} \nonumber \\
  &&\hskip 2em (\det B)^{-1}\, \delta(v\cdot A_v^{\perp +})\,
     \delta(v\cdot {A_v^{\perp +}}^\dagger)\,
     \D\Avperpl{\mu}\, \D (\Avperpl{\mu})^\dagger, \endeqa
with
\eqa B &=& i B_1\, B_2\, B_3, \\
 B_1 &=& (\Dperp)^2 +(m_V)^2, \\
 (B_2)_{\mu\nu}
 &=&[2m_V(2m_V +i\vD) +(\Dperp)^2]g_{\mu\nu} -\Dperp_\nu \Dperp_\mu \nonumber\\
   &&-(\vD -im_V)\Dperp_\mu [(\Dperp)^2 +(m_V)^2]^{-1} \Dperp_\nu
      (\vD -im_V), \\
 B_3 &=& (B_2^{-1})_{\mu\nu}\, v^\mu v^\nu, \endeqa
and
\eqa L_{\rm HVEFT}^v
   &=&(\Avperpl{\mu})^\dagger \{[-2m_V\,i\vD +(\Dperp)^2]g_{\mu\nu}
       -\Dperp_\nu \Dperp_\mu \nonumber \\
    &&\hskip 1.5cm -(\vD -im_V)\Dperp_\mu [(\Dperp)^2 +(m_V)^2]^{-1} \Dperp_\nu
        (\vD -im_V)\}\Avperpl{\nu} \nonumber \\
   &&-\{(A_{\mu,v}^{\perp +})^\dagger (\Dperp)^2
         -(\Avperpl{\alpha})^\dagger \Dperp_\mu \Dperp_\alpha \nonumber \\
   &&\hskip 1.5em -(\Avperpl{\alpha})^\dagger
          (\vD -im_V)\Dperp_\alpha [(\Dperp)^2 +(m_V)^2]^{-1} \Dperp_\nu
        (\vD -im_V)\} (B_2^{-1})^{\mu\nu} \nonumber \\
   &&\hskip 1em \{(\Dperp)^2 A_{\nu,v}^{\perp +}
         -\Dperp_\alpha \Dperp_\nu \Avperpl{\alpha} \nonumber\\
   &&\hskip 1.5em -(\vD -im_V)\Dperp_\nu [(\Dperp)^2 +(m_V)^2]^{-1} 
          \Dperp_\alpha (\vD -im_V) \Avperpl{\alpha} \}, \endeqa
Expanding out the non-local expressions in powers of $\LmdSI/m_V$ finally 
yields the heavy vector effective field theory (HVEFT) Lagrangian:
\eqa
 \Lagr_{\rm HVEFT}^v
  &=&(\Avperpl{\mu})^\dagger
     \left\{ g_{\mu\nu}\left[-2m_V\,i\vD +(\Dperp)^2 \right]
      -\Dperp_\nu \Dperp_\mu +\Dperp_\mu \Dperp_\nu \right.\nonumber \\
   &&\hskip 4em +\frac{i}{m_V} \left(\Dperp_\mu \Dperp_\nu \vD
        +\vD \Dperp_\mu \Dperp_\nu \right) \nonumber \\
   &&\hskip 4em +\frac{1}{4(m_V)^2} 
      \left[g_{\mu\nu}\left(-(\Dperp)^4 +{(\Dperp)^2\,i\vD(\Dperp)^2\over 2m_V}
       \right) +4\Dperp_\mu (\Dperp)^2 \Dperp_\nu \right.
      \label{LHVEFT1}\\
   &&\hskip 8.5em -\Dperp_\mu \Dperp_\nu (\Dperp)^2 
       -(\Dperp)^2 \Dperp_\mu \Dperp_\nu +\Dperp_\nu \Dperp_\mu (\Dperp)^2
       \nonumber\\
   &&\hskip 8.5em \left.+(\Dperp)^2 \Dperp_\nu \Dperp_\mu
       -4(\vD)\Dperp_\mu \Dperp_\nu (\vD) \right]
      +{\cal O}\left(\frac{1}{(m_V)^3}\right) \Biggl\}
     \Avperpl{\nu} \nonumber
\endeqa
Using \eqr{Aperpdef}, $\Lagr_{\rm HVEFT}$ can be rewritten in terms of 
$A_v^{\mu +}$:
\eqa
 \Lagr_{\rm HVEFT}^v
 &=&(A_v^{\mu +})^\dagger \Biggl\{ \left[ -2m_V\,i\vD +(\Dperp)^2
      -\frac{(\Dperp)^4}{4(m_V)^2}
      +\frac{(\Dperp)^2 \, i\vD(\Dperp)^2}{8(m_V)^3}\right]
       (g^{\mu\nu} -v^\mu v^\nu) \nonumber \\
 &&\hskip 4em +\Dperp_\mu \Dperp_\nu -\Dperp_\nu \Dperp_\mu
     +\frac{i}{m_V} \left(\Dperp_\mu \Dperp_\nu \vD
       +\vD \Dperp_\mu \Dperp_\nu \right) \label{LHVEFT2}\\
 &&\hskip 4em +\frac{1}{4(m_V)^2} 
      \left[g_{\mu\nu}\left(-(\Dperp)^4 +{(\Dperp)^2\,i\vD(\Dperp)^2\over 2m_V}
       \right) +4\Dperp_\mu (\Dperp)^2 \Dperp_\nu \right.
      \nonumber\\
 &&\hskip 8.5em -\Dperp_\mu \Dperp_\nu (\Dperp)^2 
       -(\Dperp)^2 \Dperp_\mu \Dperp_\nu +\Dperp_\nu \Dperp_\mu (\Dperp)^2
       \nonumber\\
 &&\hskip 8.5em \left.+(\Dperp)^2 \Dperp_\nu \Dperp_\mu
       -4(\vD)\Dperp_\mu \Dperp_\nu (\vD) \right]
     +{\cal O}\left(\frac{1}{(m_V)^3}\right) \Biggr\}
    A_v^{\nu +} \nonumber
 \endeqa

By scaling the field as
\eq A_v^{\mu +} = {A_v^{'\mu +} \over \sqrt{2m_V}}, \endeq
it can be seen from \eqr{LHVEFT2} that a theory with $N_V$ flavours 
of heavy vector particles will have a SU($3N_V$) spin-flavour symmetry
at leading order. And just as in the cases examined above, a 
Lorentz-covariant theory may be recovered by appropriately including 
the contributions from the different possible velocities. 

The remarks made in the previous investigations regarding radiative 
contributions, regularization, renormalization, and matching are also 
appropriate here. Moreover, there is a remarkable similarity between this 
analysis and the previous one involving bosons namely for spin-0 particles.

\section{Summary}

In this paper, a functional integral method for deriving an effective field 
theories for heavy particles of different spin has been presented. It gives 
the effective Lagrangian to all orders in an inverse heavy mass expansion. 
Radiative contributions can be systematically incorporated through matching 
and renormalization group running. These effective theories provide a 
convenient description of phenomena occurring below the heavy mass and in 
the kinematic region where all other interaction scales are much smaller. 
The results derived here will be utilized in the analysis of a hidden 
symmetry of heavy particle effective field theories \cite{Lee97b}.

\vskip 1cm
\leftline{\large\bf Acknowledgements} \bigskip
The author thanks B. Grinstein and A. Kapustin for fruitful discussions. 
This work was supported in part by NSERC and the Department of Energy under 
contract DOE-FG03-90ER40546.

\end{document}